\documentclass[prd,aps,showpacs,tightenlines,nofootinbib,preprint,preprintnumbers,amsmath,amssymb]{revtex4}
\usepackage{graphicx}
\usepackage{amsmath}
\input{colordvi.tex}
\def\apj #1 #2 #3 {#1, ApJ, {\bf #2}, #3}
\def\apjl #1 #2 #3 {#1, ApJ, {\bf #2}, L#3}
\def\apjs #1 #2 #3 {#1, ApJS, {\bf #2}, #3}
\def\aap  #1 #2 #3 {#1, A\&A, {\bf #2}, #3}
\def\mnras #1 #2 #3 {#1, MNRAS, {\bf #2}, #3}
\def\pra #1 #2 #3 {#1, Phys.~Rev.~A., {\bf #2}, #3}
\def\prb #1 #2 #3 {#1, Phys.~Rev.~B., {\bf #2}, #3}
\def\prc #1 #2 #3 {#1, Phys.~Rev.~C., {\bf #2}, #3}
\def\prd #1 #2 #3 {#1, Phys.~Rev.~D., {\bf #2}, #3}
\def\pre #1 #2 #3 {#1, Phys.~Rev.~E., {\bf #2}, #3}
\def\prl #1 #2 #3 {#1, Phys.~Rev.~Lett., {\bf #2}, #3}
\def\plb #1 #2 #3 {#1, Phys.~Lett.~B., {\bf #2}, #3}
\def\science #1 #2 #3 {#1, Science., {\bf #2}, #3}
\def\nature #1 #2 #3 {#1, Nature., {\bf #2}, #3}
\def\nphysa #1 #2 #3 {#1, Nucl.~Phys.~A., {\bf #2}, #3}
\def\nphysb #1 #2 #3 {#1, Nucl.~Phys.~B., {\bf #2}, #3}
\def\nphysbs #1 #2 #3 {#1, Nucl.~Phys.~B.~Suppl., {\bf #2}, #3}

\def\h#1{\hbox{${}^{#1}$H}}

\def\h502{\hbox{$ h^{2}_{50}$}}

\def\nubar{\bar\nu}
\def\la{\mathrel{\mathpalette\fun <}}
\def\ga{\mathrel{\mathpalette\fun >}}
\def\fun#1#2{\lower3.6pt\vbox{\baselineskip0pt\lineskip.9pt
  \ialign{$\mathsurround=0pt#1\hfil##\hfil$\crcr#2\crcr\sim\crcr}}}
\newcommand{\bear}{\begin{array}}  \newcommand{\eear}{\end{array}}
\newcommand{\bea}{\begin{eqnarray}}  \newcommand{\eea}{\end{eqnarray}}
\newcommand{\beq}{\begin{equation}}  \newcommand{\eeq}{\end{equation}}
\newcommand{\bef}{\begin{figure}}  \newcommand{\eef}{\end{figure}}
\newcommand{\bec}{\begin{center}}  \newcommand{\eec}{\end{center}}
\newcommand{\non}{\nonumber}

\newcommand{\bib}{\bibitem} 
\def\IBB#1#2#3{{\bf #1}, #2 (20#3)}
\def\IBID#1#2#3{{\it ibid}. {\bf #1}, #2 (19#3)}

\def\APP#1#2#3{Ann. Phys. {\bf #1}, #2 (20#3)}

\def\APJSS#1#2#3{Astrophys. J., Suppl. Ser. {\bf #1}, #2 (20#3)}

\def\JHEP#1#2#3{J. High Energy Phys. {\bf #1}, #2 (19#3)}

\def\MPLA#1#2#3{Mod. Phys. Lett. A {\bf #1}, #2 (19#3)}

\def\NPB#1#2#3{Nucl. Phys. {\bf B#1}, #2 (19#3)}
\def\NPBB#1#2#3{Nucl. Phys. {\bf B#1}, #2 (20#3)}

\def\PLB#1#2#3{Phys. Lett. B {\bf #1}, #2 (19#3)}

\def\PRD#1#2#3{Phys. Rev. D {\bf #1}, #2 (19#3)}
\def\PRDD#1#2#3{Phys. Rev. D {\bf #1}, #2 (20#3)}

\def\PRLL#1#2#3{Phys. Rev. Lett. {\bf#1}, #2 (20#3)}

\def\PRTT#1#2#3{Phys. Rep. {\bf#1}, #2 (20#3)}

\begin{document}
%\draft
%
\title{Lepton asymmetry in the primordial gravitational wave spectrum}
\author{\vspace{0.5cm} Kiyotomo Ichiki$^{1}$\footnote{E-mail address:
ichiki@resceu.s.u-tokyo.ac.jp}, Masahide Yamaguchi$^{2}$\footnote{E-mail
address: gucci@phys.aoyama.ac.jp}, and Jun'Ichi
Yokoyama$^{1,3}$
\vspace{0.5cm}}
\affiliation{ $^{1}$Research Center for the Early Universe (RESCEU),
Graduate School of Science, The University of Tokyo, Tokyo 113-0033,
Japan, \vspace{0.3cm} \\ 
$^{2}$Department of Physics and Mathematics, Aoyama Gakuin University,
Sagamihara 229-8558, Japan\vspace{0.3cm} \\
$^{3}$Galileo Galilei Institute for Theoretical Physics, Arcetri,
Firenze, Italy 
\vspace{2cm}
}
\date{\today}
\preprint{RESCEU-27/06}
\begin{abstract}
Effects of neutrino free streaming is evaluated on the 
primordial spectrum of gravitational radiation taking both
neutrino chemical potential and masses into account.
The former or the lepton asymmetry induces two competitive
effects, namely, to increase anisotropic pressure, which
damps the gravitational wave more, and to delay the
matter-radiation equality time, which reduces the damping.
The latter effect is more prominent and a large lepton
asymmetry would reduce the damping.  We may thereby be
able to measure the magnitude of lepton asymmetry
from the primordial gravitational wave spectrum.
\end{abstract}
\pacs{04.30.-w, 11.30.Fs, 98.80.-k}
\maketitle

%--------------------------------------------------------------------------
\section{Introduction}
%--------------------------------------------------------------------------
Cosmological baryon asymmetry and lepton asymmetry are among the
most fundamental
quantities of our Universe. The magnitude of baryon asymmetry is
determined by two independent methods. Baryon asymmetry can be estimated
by comparing the prediction of abundances of the light elements with
observations of those quantities in big bang nucleosynthesis (BBN)
\cite{BBN}. The observations of small scale anisotropies of the cosmic
microwave background (CMB) also determine the baryon asymmetry. Recent
results of the Wilkinson Microwave Anisotropy Probe (WMAP) coincide with the
results of BBN within two sigma level \cite{WMAP,WMAP3}. Thus,
cosmological baryon asymmetry has been determined very precisely.

%--------------------------------------------------------------------------
On the other hand, lepton asymmetry has still large uncertainties both
theoretically and observationally. At very high temperature, sphaleron
processes are active, to convert lepton asymmetry to baryon asymmetry
and vice versa \cite{sphaleron}. Then, it is naively expected that the
magnitudes of baryon and lepton asymmetry are of the same order through
these effects. But, it does not always hold true and many scenarios
which allow large lepton asymmetry have been proposed
\cite{oscillation,second,MRM,KTY,sp}. Then, the determination of the
magnitude of lepton asymmetry is crucial to determine cosmic history of
our universe.  Contrary to the case of
 baryon asymmetry, however, cosmological
lepton asymmetry cannot be observed directly and only upper bounds on
lepton asymmetry has been
 obtained by indirect methods such as BBN, CMB, large
scale structure, and so on. The most stringent constraint comes from BBN
and the degeneracy parameter $\xi$ %,namely the ratio of the chemical
%potential $\mu$ to the (effective) temperature, 
is constrained as
\beq
  |\xi_{\nu_e}| < 0.2 \quad {\rm or} \quad |\xi_{\nu_\mu,\nu_\tau}|
 < 2.6
\eeq 
for the case of no mixing \cite{HMMMP}\footnote{
In case that the solution of the solar neutrino problem is other than
the large mixing angle MSW solution and that the angle $\theta_{13}$
is sufficiently small but non-zero, flavor mixing can be suppressed
\cite{smixing}. Furthermore,  the introduction of hypothetical
coupling of neutrino can also suppress flavor mixing  \cite{nomixing}.
} and
\beq
  |\xi| < 0.07
\eeq 
for each family in the case of strong mixing \cite{smixing}.
Here the degeneracy parameter $\xi$ is the ratio of the chemical
potential $\mu$ to the (effective) temperature,
\beq
\xi = \frac{\mu}{T_\nu}~.
\eeq
%

%--------------------------------------------------------------------------
In this paper, we argue that lepton asymmetry may be
 directly measurable
if one can observationally determine 
the primordial gravitational wave spectrum precisely.
During inflation, stochastic gravitational waves are generated and
expanded to cosmological scales \cite{staro}. 
Then, such gravitational waves memorize
all cosmic history during and after inflation, and provide us with
fruitful information. For example, the amplitude of the spectrum gives
us the energy scale of inflation directly. The spectral shape reflects
the history of cosmic expansion: the spectrum is roughly proportional to
$f^{-2}$, $f^0$, $f$ for the modes which reenter the horizon in the
matter dominated phase, the radiation dominated phase, and the
kinetic-energy dominated phase. 
This feature may be used to probe the possible change of the equation
of state in the early universe \cite{seto}.
Recently, Watanabe and Komatsu took into
account the change of the effective number of degrees of freedom of all
standard model particles of particle physics and showed that such
changes leave characteristic features in the spectrum
\cite{WK}.\footnote{The effect of quark gluon plasma phase transition is
discussed in \cite{QGP}.} Furthermore, Weinberg pointed out that the
square amplitude of the gravitational wave is reduced by 35.6\% if
neutrino free streaming effects are taken into account \cite{weinberg}.

%--------------------------------------------------------------------------
In this paper, we evaluate neutrino free streaming effects on the
primordial gravitational wave spectrum by taking into account lepton
asymmetry (chemical potential of neutrinos) and neutrino masses. Lepton
asymmetry induces two opposite effects. One is to increase 
anisotropic pressure, which increases the damping of the spectrum by
free streaming effects of neutrinos. The other is to delay the 
matter-radiation equality, which decreases it. Then, we show that net
effect of lepton asymmetry is to decrease
 the damping of the amplitude of the
primordial gravitational wave by free streaming effects of neutrinos.

%--------------------------------------------------------------------------
In the next section, we briefly review the basics of lepton asymmetry
and derive anisotropic pressure of neutrino in the presence of lepton
asymmetry. In Sec. III, we investigate the evolution of gravitational
wave and discuss the damping of the amplitude of the primordial
gravitational wave by free streaming effects of neutrinos. Finally we
give our discussion and summary.

%--------------------------------------------------------------------------
\section{Lepton Asymmetric Cosmology}
%--------------------------------------------------------------------------

In this section we quickly review the basics of massive degenerate
neutrinos. Observational implications of degenerate massive neutrinos
for density perturbations have been
 discussed in the literature \cite{LP,OKMW,LRV}.
Here we apply them to the evolution of primordial gravitational wave
background. 

%--------------------------------------------------------------------------
\subsection{background}

The energy density of one species of massive degenerate neutrinos and
anti-neutrinos is given by \cite{FKT,LP}
\begin{eqnarray}
\rho_\nu+\rho_{\bar\nu}&=&a^{-4}(k_B
 T_{\nu,0})^4\int\frac{d^3q}{(2\pi)^3}\sqrt{q^2+a^2m^2}
 \left(f_\nu(q)+f_{\bar\nu}(q)\right)\nonumber \\
 &=& (k_B T_{\nu})^4
 \int\frac{d^3q}{(2\pi)^3} \epsilon \left(f_\nu(q)+f_{\bar\nu}(q)\right)~.
\end{eqnarray}
Here $T_\nu$ and $T_{\nu_0}$ are the temperature of neutrinos and that
of today, respectively, $a$ is a cosmic scale factor, 
$q$ is comoving momentum in units of $k_B T_{\nu,0}$, $m$ is
normalized neutrino mass defined by 
\begin{equation}
m=\frac{m_\nu}{kT_{\nu,0}}~,
\label{Eq:mass_param}
\end{equation}
and $\epsilon\equiv \sqrt{q^2+a^2m^2}$ is the comoving energy
density. Note that comoving momentum $q$ is related to the proper
momentum $p$ by $p = q/a$, which redshifts as $a^{-1}$. 
Therefore, $q$ is constant relative to the expansion of the universe.
When the universe was dense and hot enough, neutrinos were kept
in thermal equilibrium with the rest of the plasma through weak
interactions so that the distribution function had been
relaxed to the 
form of Fermi-Dirac distribution. 
Even after neutrino decoupling at $\sim$MeV, owing to Liouville's
theorem, the distribution function is still given by the distribution: 
\begin{equation}
 f_\nu(q)=\frac{1}{1+e^{q-\xi}}~,\hspace{1cm}
 f_{\bar\nu}(q)=\frac{1}{1+e^{q+\xi}}~.
\end{equation}
Here $\xi$ is the degeneracy parameter.

%--------------------------------------------------------------------------
The pressure is given by
\begin{equation}
3\left(P_\nu+P_{\bar\nu}\right)=(k_B T_\nu)^4
\int\frac{d^3q}{(2\pi)^3}\frac{q^2}{\epsilon} 
\left(f_\nu(q)+f_{\bar\nu}(q)\right)~.
\end{equation}
Here and hereafter, we assume that $\mu$ (and $\xi$) is spatially
homogeneous for simplicity. 
In our calculation we also assume, for simplicity, that all the three
generations have the same properties, that is,
 we set $m_{\nu_e}=m_{\nu_\mu}=m_{\nu_\tau}\equiv m_\nu$  and
$\xi_{\nu_e}=\xi_{\nu_\mu}=\xi_{\nu_\tau}\equiv \xi$.

%--------------------------------------------------------------------------
It will be useful to note the expressions for the energy density and the
pressure of massive degenerate neutrinos, in the relativistic and
non-relativistic limits for numerical stability and efficiency.
In the relativistic limit, the energy density can be written as
\begin{eqnarray}
\rho_\nu+\rho_{\bar\nu}
&\approx&\frac{(k_B T_\nu)^4}{2\pi^2}\left[\int q^3 dq
				      (f_\nu+f_{\nubar})+\int q^3
				      dq(f_\nu+f_{\nubar})
				      \frac{(am)^2}{2q^2}\right]
\nonumber \\
&=&(\rho_{\nu_0}+\rho_{\bar\nu_0})\left[\left\{1+\frac{30}{7}
\left(\frac{\xi}{\pi}\right)^2+\frac{15}{7}\left(\frac{\xi}{\pi}\right)^4\right\}
 +\left(\frac{5}{7\pi^2}\right)\left\{1+3\left(\frac{\xi}{\pi}\right)^2
			       \right\}a^2m^2\right]~.
\label{eq:ED_rel_limit}
\end{eqnarray}
Here we have used the well-known expression for the energy density of
massless non-degenerate neutrinos
\begin{equation}
\rho_{\nu_0}+\rho_{\bar\nu_0} = (k_B T_\nu)^4\int\frac{d^3q}{(2\pi)^3}q
 \frac{2}{1+e^q} = \frac{2(k_B T_\nu)^4}{2\pi^2}\left(\frac{7\pi^4}{120}\right)~.
\end{equation}
In the same manner, the expression for the pressure can be found to be
\begin{equation}
 3(P_\nu+P_{\nubar})\approx 2(\rho_{\nu_0}+\rho_{\bar\nu_0})
\left\{1+\frac{30}{7}\left(\frac{\xi}{\pi}\right)^2+\frac{15}{7}\left(\frac{\xi}{\pi}\right)^4\right\}
-(\rho_\nu+\rho_{\bar\nu})~.
\end{equation}

%----------------------------------------------------------------------
In the non-relativistic limit, on the other hand, we shall expand
$\epsilon$ as $\epsilon \approx
am(1+\frac{q^2}{2a^2m^2})$. Unfortunately, we do not have any expression
for the even moments of degenerate distribution function in terms of a
finite power series of $\xi$. Therefore we keep the integral expressions
and write the energy density as
\begin{eqnarray}
\rho_\nu+\rho_{\nubar}&\approx& \frac{(k_BT_\nu)^4}{2\pi^2}
 \left[am \int q^2 dq (f_\nu+f_{\nubar})+\frac{1}{2am}\int q^4 dq
  (f_\nu+f_{\nubar})\right]\nonumber \\
&=&(\rho_{\nu_0}+\rho_{\nubar_0})\left(am\Theta_2+\frac{1}{2am}\Theta_4\right)~, 
\end{eqnarray}
and the pressure as
\begin{eqnarray}
3(P_\nu+P_{\nubar})&\approx& \frac{(k_BT_\nu)^4}{2\pi^2}\left[
\frac{1}{am}\int q^4 dq(f_\nu+f_{\nubar})-\frac{1}{2(am)^3}\int q^6 dq (f_\nu+f_{\nubar})
\right] \nonumber \\
&=&(\rho_{\nu_0}+\rho_{\nubar_0})\left(\frac{1}{am}\Theta_4-\frac{1}{2(am)^3}\Theta_6\right)~.
\end{eqnarray}
Here we have defined the even-moments of distribution function
$\Theta_n$, normalized by the third moment of non-degenerate fermion
distribution, $\int q^3 dq \frac{2}{(1+e^q)}=2(\frac{7\pi^4}{120})$, as
\begin{equation}
\Theta_n(\xi) = \frac{1}{2}\int q^n dq (f_\nu+f_{\nubar})\left/\left(\frac{7\pi^4}{120}\right)\right.~.
\end{equation}

%---------------------------------------------------------------------
Before finishing this subsection we would like to comment on the
relation between the cosmological neutrino density parameter
$\Omega_\nu$ and the mass of a neutrino $m_\nu$ \cite{FKT}.
\begin{figure}
  \rotatebox{0}{\includegraphics[width=0.5\textwidth]{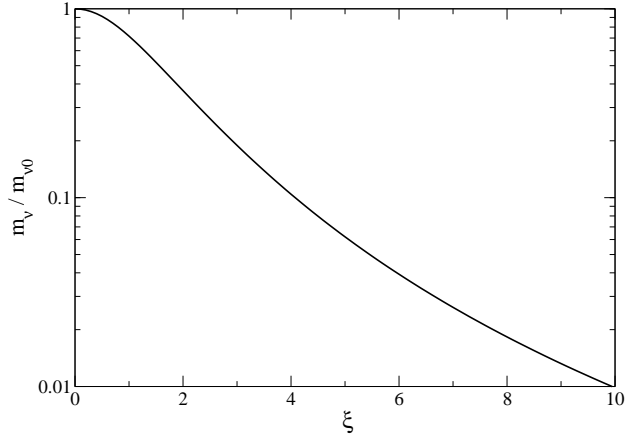}}
\caption{The neutrino mass as a function of $\xi$ normalized by $\xi=0$
 non-degenerate  neutrino. With $\Omega_\nu h^2$ fixed, the mass of 
 neutrinos in lepton  asymmetric models are lighter than
 that in symmetric models. }
\label{fig:mass_ratio}
\end{figure}
Basically, as long as $m_\nu$ is larger than $k_BT_\nu$,
the neutrino mass is related to $\Omega_\nu$ as
\begin{equation}
m_\nu = \frac{\Omega_\nu \rho_{\rm crit}}{N_\nu (n_\nu+n_{\nubar})}~,
\end{equation}
where $\Omega_\nu$ is the density parameter of neutrinos, $\rho_{\rm
crit}$ is the critical density, $N_\nu$ is the number of species of
massive neutrino, $n_\nu$ is the number density of one species of massive
neutrino. In the case of degenerate neutrino, the number density is
different from that of non-degenerate one, and it is given by
\begin{eqnarray}
(n_\nu+n_{\nubar})&=&(k_B T_\nu)^3\int \frac{d^3
 q}{(2\pi)^3}(f_\nu(q)+\bar{f}_{\nubar}(q))\nonumber \\
&=&\frac{(k_B T_\nu)^3}{2\pi^2}\times 2
 \left(\frac{7\pi^4}{120}\right) \Theta_2~.
\label{eq:n_mu}
\end{eqnarray}
By remembering that for non-degenerate massive neutrinos the number
density is expressed as
\begin{equation}
(n_{\nu_0}+n_{\nubar_0})=\frac{(k_B
			  T_\nu)^3}{2\pi^2}\times
 2\times\frac{3}{2}\zeta(3)~,
\end{equation}
where $\zeta$ is the zeta function, we obtain the
following relation between masses of degenerate neutrino and
non-degenerate one as  
\begin{equation}
\sum m_\nu = \frac{\frac{3}{2}\zeta(3)}{\frac{7\pi^4}{120}
 \Theta_2(\xi)}\sum m_{\nu_0}\left(\frac{T_{\nu}(0)}{T_\nu(\xi)}\right)^3~.
\label{eq:sum_m}
\end{equation}
Here $\sum m_{\nu_0}=m_{{\nu_e}_0}+m_{{\nu_\mu}_0}+m_{{\nu_\tau}_0}=94.1$ eV $\Omega_{\nu}h^2$ is the neutrino mass 
in the non-degenerate cases, 
and $T_\nu(0)$ and $T_{\nu}(\xi)$ are
the present temperatures of standard and degenerate neutrinos,
respectively.

The temperatures of degenerate and non-degenerate neutrinos can differ
greatly from each other if the chemical potential of neutrinos is 
so large that neutrinos decouple from the
rest of the plasma before muon-antimuon annihilation. In reality,
however, neutrinos can be kept in kinetic
equilibrium with $e^\pm$ even with a significant degeneracy to hold the
same temperature as the rest of the plasma until at least the
muon-antimuon annihilation ends \cite{IK2003}, 
while number freeze-out can occur before cosmological QCD phase
transition \cite{KS1992,OKMW2}. 
So, we can safely set 
$T_\nu(0)=T_\nu(\xi)=\left(\frac{4}{11}\right)^{1/3}T_{\gamma,0}$ in
Eq.(\ref{eq:sum_m}).
Therefore, with the density parameter
$\Omega_\nu h^2$ being fixed, the mass should
be monotonically smaller as the neutrinos are degenerate larger. 
The mass ratio between degenerate and non-degenerate cases is plotted in
Fig.\ \ref{fig:mass_ratio}.

%---------------------------------------------------------------------
\subsection{perturbations}

In this paper we only consider the tensor type perturbations. The line
element is given by
\begin{equation}
ds^2 = g_{\mu\nu}dx^\mu dx^\nu=a^2(\tau)\left[-d\tau^2+(\delta_{ij}+h^T_{ij})dx^i dx^j\right]~,
\end{equation}
where $h^T_{ij}$ denotes tensor type perturbation (gravitational waves)
around flat Friedmann-Robertson-Walker metric. We shall work in
the transverse traceless gauge, which is set by the conditions
$h^T_{ij,i}=h^T_{ii}=0$. 
The evolution of gravitational waves $h^T_{ij}$ is determined by the
linearized Einstein equations:
\begin{equation}
\ddot{h}^{T}_{ij}+2\left(\frac{\dot{a}}{a}\right)\dot{h}^T_{ij}+k^2
 h^T_{ij}=16\pi G a^2 \pi_{ij}~,
\label{eq:evol_gw}
\end{equation}
where dots indicate conformal time derivatives, and $\pi_{ij}$ is the
anisotropic part of the stress tensor $T_{ij}$, defined by
\begin{equation}
a^2 \pi_{ij}=T_{ij}-pg_{ij}~.
\end{equation}

Following the standard procedure, we expand the distribution function of
neutrinos into homogeneous and perturbed inhomogeneous parts as
\begin{equation}
 f_\nu+f_{\nubar}=(\bar{f}_\nu+\bar{f}_{\nubar})(1+\Psi^T(x^i,\tau,q_i))~.
\end{equation}
Note that we take comoving momentum in the tetrad frame $q_i$ as our
momentum variable, which is related to the (spatial part of) four
momentum $P^i$ as $q_i = a^2 P^j(\delta_{ij}+\frac{1}{2}h_{ij})$. 
By virtue of this momentum variable, one can express the moments of
distribution function such as energy density, pressure and anisotropic
stress in a simple manner, {\it i.e.},
 $q_i$ can be treated as that on the flat
space background.
For example, the anisotropic stress of neutrinos can be written as
\begin{equation}
\pi^T_{ij} = (kT_\nu)^4\int \frac{d^3 q}{(2\pi)^3}\frac{q^2}{\epsilon}(\gamma_i
 \gamma_j-\frac{1}{3}\delta_{ij})(f_\nu+f_{\nubar})\Psi^{T}
\label{eq:def_pi}
\end{equation}
where we write the comoving momentum $q_i$ in terms of its magnitude and
direction: $q_i=q \gamma_i$, with $\delta^{ij}\gamma_i\gamma_j=1$.

The perturbed distribution function $\Psi^T$ satisfies tensor type
collisionless Boltzmann equation in $k$-space 
\begin{equation}
 \frac{\partial \Psi^T_k}{\partial \tau}+i\frac{q}{\epsilon}
k\mu\Psi^T_k -\frac{1}{2}\dot{h}_{ij}\gamma^i\gamma^j\frac{d\ln
  (\bar{f}_\nu+\bar{f}_{\nubar})}{d\ln q}=0~,
\label{eq:Boltz_tensor}
\end{equation}
where $\mu$ is a directional cosine between $\vec{k}$ and $\vec{q}$, 
and
logarithmic derivative of the distribution function with respect to
$q$ is given by
\begin{equation}
\frac{d\ln (\bar{f}_\nu+\bar{f}_{\nubar})}{d\ln q} =
 -\frac{q}{e^{-q}+\cosh\xi}\left(\frac{1+\cosh\xi\cosh q}{\cosh\xi+\cosh q}\right)~.
\label{eq:q-dependence}
\end{equation}
One finds the solution of Eq.(\ref{eq:Boltz_tensor}) as
\begin{equation}
\Psi^T = \frac{1}{2}\frac{d\ln (\bar{f}_\nu+\bar{f}_{\nubar})}{d\ln
 q}\gamma^i \gamma^j \int^u_0 du^\prime e^{i\mu
 (u^\prime-u)}h^\prime_{ij}(u^\prime)~,
\label{eq:sol_tensor}
\end{equation}
where new time variable $u$ is introduced by
\begin{equation}
 u = u(q)\equiv k\int^\tau_{\tau_{\rm
  dec}}\frac{d\tau^\prime}{\sqrt{1+\frac{a^2 m^2}{q^2}}}~.
\label{eq:def_u}
\end{equation}
One of the effects of finite mass comes through Eq.(\ref{eq:def_u}),
which delays the timing of new time variable $u$ compared  with the
conformal time $k\tau$.

%--------------------------------------------------------------
By inserting the solution Eq.(\ref{eq:sol_tensor}) into
Eq.(\ref{eq:def_pi}), and integrating by parts, one obtains
\begin{equation}
\pi^T_{ij}=-\frac{(k_B T_\nu)^4}{16}\int \frac{q^2 dq}{2\pi^2}(\bar{f}_\nu+\bar{f}_{\nubar})\frac{q^2}{\epsilon}\left(5-\frac{q^2}{\epsilon^2}\right)\int^{u(q)}_0 du^\prime \int^{1}_{-1}dx
 e^{ix(u-u^\prime)}(1-x^2)^2\frac{\partial }{\partial
 u^\prime}h_{ij}(u^\prime)
\label{eq:sol_pi}
\end{equation}
One can easily confirm that, in cases of massless non-degenerate
neutrinos, Eq.(\ref{eq:sol_pi}) reduces to the results obtained in
\cite{weinberg,WK}. By substituting this expression into the l.h.s. of
Eq.(\ref{eq:evol_gw}), one sees that the evolution of gravitational waves
is given by solving an integro-differential equation.

%--------------------------------------------------------------------
In order to see how neutrino asymmetry affects the evolution of
gravitational waves, let us first consider massless degenerate neutrinos.
When $m=0$, the expression for $\pi^T_{ij}$ reduces to
\begin{equation}
\pi^T_{ij, \mbox{\scriptsize
 massless}}=-4(\rho_{\nu}+\rho_{\nubar})\int^u_0 du^\prime\int^1_{-1}dx
\frac{(1-x^2)^2}{16}e^{ix(u-u^\prime)}\frac{\partial}{\partial
 u^\prime}h^T_{ij}(u^\prime)~.
\label{eq:pi_massless}
\end{equation}
In such cases, therefore, $q$ dependence of the distribution functions
(Eq. (\ref{eq:q-dependence})) can be integrated away into the energy
density, just like in the non-degenerate cases. So the effect of $\xi$
arises only through the background energy density (in the first term of
Eq. (\ref{eq:ED_rel_limit})), and is completely described by introducing
an effective number of massless non-degenerate neutrinos. 

For massive
degenerate neutrinos, however, the situation is quite different. One
can no longer express the momentum integration in Eq. (\ref{eq:sol_pi})
in terms of well-known moments such as density, pressure and so on.
In other words, it will not be possible to renormalize the dependence of
$\xi$ by the effective number of massless non-degenerate
neutrinos. Thus, in principle, these two contributions are
distinguishable. 

In the relativistic limit, the expression of $\pi^T_{ij}$ can be
expanded to 
\begin{eqnarray}
\pi^T_{ij}&\approx& -(k_B T_\nu)^4\int \frac{q^2 dq}{2\pi^2}(\bar{f}_\nu+\bar{f}_{\nubar})q
      \left(4-\frac{a^2 m^2}{q^2}\right) \non \\
 && \qquad \qquad \qquad \times 
\int^u_0 du^\prime\int^{1}_{-1}dx
 e^{ix(u-u^\prime)}\frac{(1-x^2)^2}{16}\frac{\partial }{\partial
 u^\prime}h_{ij}(u^\prime)\nonumber \\
&=&\pi^T_{ij, \mbox{\scriptsize massless}}+a^2 m^2
 (\rho_{\nu_0}+\rho_{\nubar_0})\left(\frac{10}{7\pi^2}\right)
 \left\{1+3\left(\frac{\xi}{\pi}\right)^2\right\} 
 \non \\
&& \qquad \qquad \qquad \times
\int^u_0 du^\prime\int^{1}_{-1}dx
 e^{ix(u-u^\prime)}\frac{(1-x^2)^2}{16}\frac{\partial }{\partial
 u^\prime}h_{ij}(u^\prime)~.
\label{eq:sol_pi_rel_limit}
\end{eqnarray}
In the non-relativistic limit, on the other hand, $\pi^T_{ij}$ can be
expressed as
\begin{equation}
\pi^T_{ij}\approx  -\frac{5}{am}\Theta_4(\xi)(\rho_{\nu_0}+\rho_{\nubar_0})\int^u_0 du^\prime\int^{1}_{-1}dx
 e^{ix(u-u^\prime)}\frac{(1-x^2)^2}{16}\frac{\partial }{\partial
 u^\prime}h_{ij}(u^\prime)
\end{equation}
For both limits, a finite mass always acts to suppress the anisotropic
stress relative to the massless case.
One finds that when neutrinos become massive, $q/\epsilon \to 0$, then
$\pi^T_{ij}\to 0$.

%--------------------------------------------------------------------------
\section{evolution of gravitational waves}
%--------------------------------------------------------------------------

In the absence of neutrino free streaming
 in which $\pi^T_{ij}=0$, the evolution equation of
gravitational waves (Eq. (\ref{eq:evol_gw})) with an appropriate initial
condition has well-known analytical solutions:  $h \propto j_0(k\tau)$
in the radiation dominated era, and $h\propto j_1(k\tau)/(k\tau)$ in the
matter dominated era. Thus for both eras, the amplitude of gravitational
waves diminishes as $a^{-1}$ with oscillations when waves are well
inside the cosmic horizon ($k\tau\gg 1$). This is a general feature of
waves in an expanding universe.

\begin{figure}
  \rotatebox{0}{\includegraphics[width=0.5\textwidth]{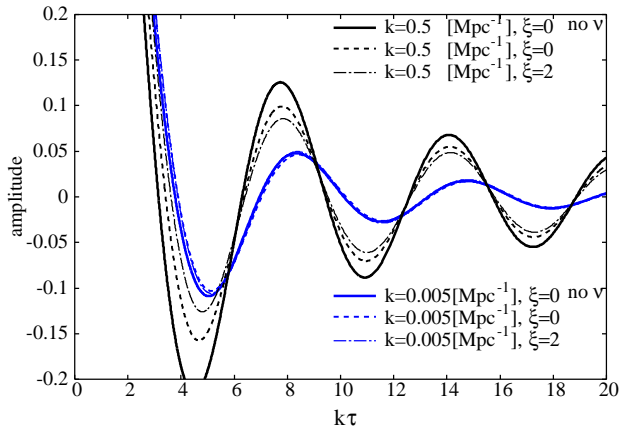}}
\caption{The effects of neutrino free streaming and $\xi$ in the time
 evolution of gravitational waves. The black lines show the numerical
 solutions of gravitational waves with wavenumber $k=0.5$ Mpc$^{-1}$,
 and blue (gray) lines show those with wavenumber $k=0.005$ Mpc$^{-1}$. 
 The two thickest lines are solutions  without neutrinos (no $\nu$).
 The dash-dotted lines are numerical solutions with degeneracy
 parameter $\xi=2$. The straight thick and dashed lines are
 those without chemical potentials. The lines with smaller wavenumber
 ($k=0.005$) are almost indistinguishable in this figure.}
\label{fig:1}
\end{figure}

The main effect of neutrinos on this evolution of gravitational waves is
the damping of amplitude by their free streaming \cite{weinberg}, and it
can be significant when the universe is radiation dominated and
neutrinos are relativistic. Because the gravitational waves with
wavenumber $k\ga 0.01$ Mpc$^{-1}$ enter the Hubble horizon before
matter-radiation equality for the case of standard massless neutrinos, they
suffer from significant damping. This condition holds true unless mass
of neutrinos is heavier than $\sim 1$ eV, in which cases neutrinos
become non-relativistic before the equality. On the contrary, the
gravitational waves with longer wavenumber $k\la 0.01$ Mpc$^{-1}$ are
not significantly affected by neutrinos.  Examples of evolutions of
gravitational waves with wavenumbers $k=0.005$, $0.5$ Mpc$^{-1}$ are
illustrated in Fig.\ \ref{fig:1}.

In Fig.\ \ref{fig:1}, we also show the effect of non-zero chemical
potential on the evolution of gravitational waves. Non-zero chemical
potential modifies the evolution of gravitational waves mainly in two
ways through an effective increase of energy density of neutrinos.  One
is an effective increase of anisotropic stress in
Eq.\ (\ref{eq:pi_massless}).  This increases the effect of free streaming
neutrinos, and consequently, the damping of gravitational waves becomes
larger. The effect is clearly seen in the numerical evolution of
gravitational waves (black dash-dotted line) in Fig.\ \ref{fig:1}.  The
other is the shift of matter-radiation equality to the later time in the
history of the expanding universe. This indirectly gives rise to an
overall increase in amplitude of gravitational waves at present for
modes which have come across the horizon in the radiation dominated
universe. 

Let us quantitatively discuss the former effect first.  An analytic
estimation of the damping of gravitational waves by neutrino free
streaming was given by \cite{DR} in a sophisticated way. Here we follow
their argument and apply it to the lepton asymmetric cosmology.  They
expanded a solution of gravitational waves $h(u)$ in a series of
spherical Bessel functions as
\begin{equation}
h(u)=\sum_{n=0}^\infty a_n j_n(u)~.
\end{equation}
At sufficiently late times, the amplitude of gravitational waves
asymptotically takes of the form
\begin{equation}
h(u)=A \sin(u+\delta)/u
\end{equation}
For large argument, all of the even order Bessel functions go as $\pm
\sin x/x$ so the damping factor $A$ is given by
\begin{equation}
A \approx \sum_{n=0}^5 (-1)^n a_{2n}~.
\end{equation}
Coefficients $a_{2n}$ are determined by the recursion relation
\begin{equation}
a_{2n}=-1.6 f_\nu \frac{\sum_{k=0}^{n-1}
 a_{2k}c_{2k,2n-2k}}{2n(2n+1)+1.6f_\nu}~, 
\end{equation}
starting with $a_0=1$, where $f_\nu$ is the fraction of the energy
density in neutrinos.  Numerical coefficients $c_{2k,2n-2k}$ are listed
in \cite{DR}.

\begin{figure}
  \rotatebox{0}{\includegraphics[width=0.5\textwidth]{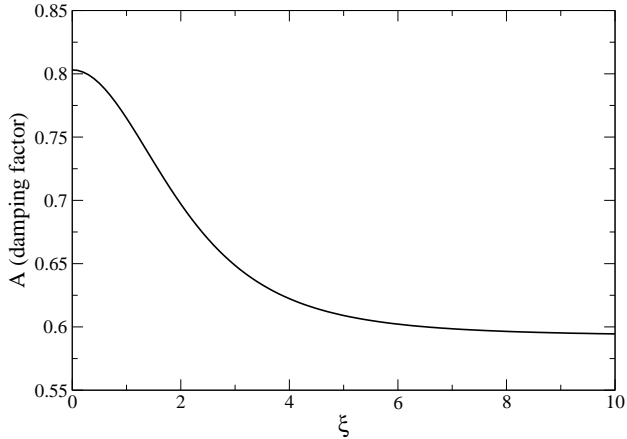}}
\caption{The damping factor as a function of degeneracy parameter
 $\xi$. The damping effect by neutrino free streaming becomes larger
 with larger $\xi$.} 
\label{fig:3}
\end{figure}

For the standard (massless and non-degenerate) neutrinos,
$f_\nu=0.40523$, which gives $A=0.80313$ \cite{DR}.  However, in the
lepton asymmetric cosmology, it is not true in general. Using degeneracy
parameter $\xi$, the fraction of the energy density in neutrinos is now
given by
\begin{equation}
f_\nu(\xi)= \frac{F(\xi)}{F(\xi)+1.46773}~,
\end{equation}
where
\begin{equation}
F(\xi)= 1+\frac{30}{7}\left(\frac{\xi}{\pi}\right)^2+\frac{15}{7}\left(\frac{\xi}{\pi}\right)^4~.
\end{equation}
When $\xi=\pi$, we get $A\approx 0.64231$. This enhancement of damping
by neutrino free streaming is illustrated in Fig.\ \ref{fig:3}.

Next, we discuss the second effect, {\it i.e.}, 
the shift of matter-radiation
equality. This can result in a modification of the spectrum of
gravitational waves at present. In order to investigate it, let us
introduce the power spectrum of gravitational waves as
\begin{equation}
\Delta_h^2 = \frac{2 k^3}{2\pi^2}\left<|h|^2\right>~.
\end{equation}
In the simplest inflationary models, it is expected that the power
spectrum is nearly scale invariant and has the amplitude of 
$\frac{16}{\pi}\left(\frac{H_{\rm inf}}{m_{\rm pl}}\right)^2$, where
$H_{\rm inf}$ is the Hubble parameter during inflation.  Alternatively,
the spectral energy density, $\Omega_h(k)\propto k^2 \Delta_h^2$, is
often used instead of the power spectrum. It has been shown that, as a
result of the evolution of gravitational waves from the deep radiation
dominated era through the matter dominated era to the present universe,
the spectral energy density has a flat spectrum at $k\ga 0.01$
Mpc$^{-1}$, where the corresponding Fourier modes have come across the
cosmic horizon during radiation dominated era, and $k^{-2}$ at $k\la
0.01$ Mpc$^{-1}$, where those have crossed the horizon during matter
dominated era \cite{Allen}. We shall discuss the spectrum of
gravitational waves in terms of this spectral energy density.

\begin{figure}
  \rotatebox{0}{\includegraphics[width=0.5\textwidth]{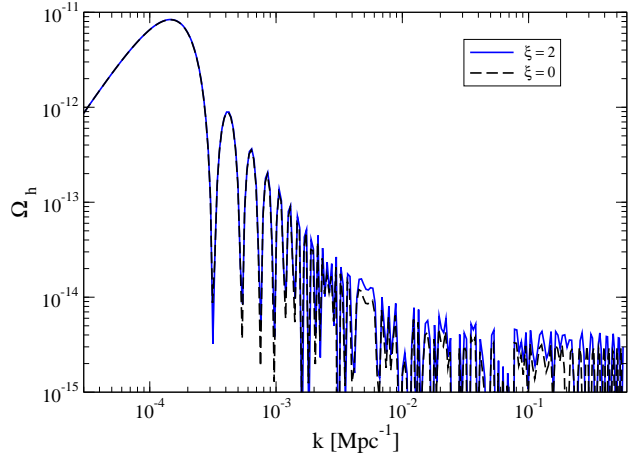}}
 \caption{Spectra of gravitational waves with and without lepton
 asymmetry. The net effect from lepton asymmetry is a rise of the plateau
 at $k > 10^{-2}$ Mpc$^{-1}$.}
\label{fig:amplitude_boost_damp}
\end{figure}

Because the energy density spectrum of gravitational waves, whose
corresponding Fourier modes cross the cosmic horizon during the matter
dominated epoch, has a slope of $k^{-2}$ if the primordial spectrum was
scale invariant, the amplitude will be boosted
by a shift of matter-radiation equality, provided that  the
normalization of overall 
amplitude of gravitational waves is given at superhorizon scales.
The amplification factor is therefore given by, 
\begin{equation}
\frac{\Omega_h}{\Omega_{h,\rm standard}}=\left(\frac{k_{\rm eq}}{k_{\rm eq, standard}}\right)^{-2}=\frac{2+1.36F(\xi)}{3.36}~.
\end{equation}
To derive this relation we have used the fact that the redshift at the
matter-radiation equality is given by
\begin{equation}
1+z_{\rm eq}=\frac{\rho_{\rm crit}\Omega_{\rm
 m}}{\rho_{\gamma,0}+\rho_{\nu,0}}
=\frac{8.83\times 10^3}{2+1.36F(\xi)}
\left(\frac{\Omega_{\rm m}h^2}{0.11}\right)~,
\end{equation}
and the corresponding comoving wavenumber is therefore
\begin{equation}
 k_{\rm eq} = \frac{H_{\rm eq}}{1+z_{\rm eq}}=0.0145 \mbox{ [Mpc$^{-1}$]}
\left(\frac{3.36}{2+1.36F(\xi)}\right)^{1/2}\left(\frac{\Omega_{\rm
	      m}h^2}{0.11}\right)~.
\end{equation}
Here $\Omega_m$ is total matter density in units of the critical density
$\rho_{\rm crit}$, and $\rho_{\gamma, 0}$, $\rho_{\nu_0}$ are energy
densities of photons and neutrinos today.
When $\xi=\pi$, $\frac{\Omega_h}{\Omega_{h, \rm st}}\approx 3.6$.  This
indirect amplification of gravitational waves dominates the damping
effect discussed earlier. This is illustrated in
Fig.\ \ref{fig:amplitude_boost_damp}.

\begin{figure}
\vspace*{1cm}
  \rotatebox{0}{\includegraphics[width=0.5\textwidth]{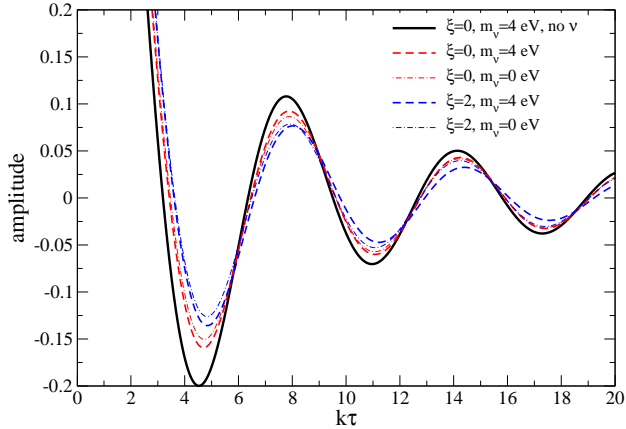}}
\caption{The effects of finite neutrino mass ($m_\nu=4$ eV) in the evolution
 of gravitational waves with the wavenumber $k=0.05$ Mpc$^{-1}$. The
 finite mass suppresses the anisotropic stress of neutrinos and the
 damping effect becomes smaller.}
\label{fig:finite_mass}
\end{figure}

As mentioned earlier, if the mass of neutrinos are heavier than $\sim 1$
eV, neutrinos have gone out from the ultra-relativistic regime before the
matter radiation equality and then made their free streaming effect on
gravitational waves smaller. In Fig.\ \ref{fig:finite_mass}, we
depict the effect of finite mass of neutrinos on the time evolution of
gravitational waves.  From a practical viewpoint, however, the mass of
neutrinos as large as $\ga 1$ eV is not realistic because even CMB
anisotropy data alone by WMAP have already put constraint on the mass as
$\sum m_\nu \la 2$ eV (at 95\% confidence) \cite{FIKL} and the
constraint can be tighter $\sum m_\nu \la 0.68$ eV if the information
from the matter power spectra is included \cite{WMAP3}. Therefore we may
conclude that the effect of finite mass of neutrinos cannot be
significant on the evolution of gravitational waves.

%--------------------------------------------------------------------------
\section{Discussion and Summary}
In this paper we investigated the evolution of gravitational wave
background in lepton asymmetric cosmology. We showed that the non-zero
degeneracy of neutrinos can modify the present spectrum of gravitational
wave background through two ways. One is that the non-negligible
chemical potential results in the larger anisotropic stress of neutrinos 
and thus larger damping of amplitude of gravitational waves.
The other is the shift of matter radiation equality, which indirectly
raises the plateau of the spectrum at $k \ga 10^{-2}$ Mpc$^{-1}$. These
effects can act in the opposite direction from each other. In
fact, we showed that the latter effect dominates the former, and the
spectrum has a larger power in total than that in the case with
non-degenerate neutrinos.
These effects follow from the
effective increase of number of massless neutrinos. We presented
straightforward numerical calculations and  analytic estimates for both
of the effects.

An important implication of our results is that, should the other 
cosmological parameters such as $\Omega_m h^2$, $n_T$, $m_\nu$ ever be
known, one can determine in principle the lepton asymmetry by measuring
the gravitational wave amplitude around $k\ga 10^{-2}$Mpc$^{-1}$. This fact
would become more hopeful noting that, among
these cosmological parameters, neutrino mass would not be so important because
the mass itself does not affect the shape of the spectrum of
gravitational waves once it is constrained to be less than $\sim$
eV from the other cosmological or terrestrial experiments.

Of course, the detection of gravitational wave background is promising
but still very
challenging. Because the damping due to neutrino free streaming can
be found only at lower frequency than $\sim 2\times 10^{-10}$ Hz
\cite{WK} 
 assuming standard decoupling of neutrinos, it will be difficult to
detect it in a direct way. Indeed, although the critical frequency would
be higher as a consequence of smaller Hubble horizon scale at neutrino
decoupling 
in the presence of a significant degeneracy, the frequency
will be still 
far lower than that aimed by the proposed gravitational wave
experiments. 
The most promising way to measure the lepton
asymmetry would be to detect the gravitational wave background
indirectly by observing the curl modes of cosmic microwave background
polarization patterns.  It is expected that, in the future planned
experiments of CMB anisotropies such as Planck or CMBpol, the lepton
asymmetry would be determined if the degeneracy parameter is as large
as $\xi \sim 0.1$ and the contamination from divergence mode of
polarization by cosmic shear is cleaned out.

As stated in the introduction, in case that flavor equilibrium between
all active neutrino species is established before BBN epoch, which is
indicated by the large mixing angle solution of the solar neutrino
problem, light
element abundances have already put a tight conservative limit on
degeneracy parameter as $|\xi| \la 0.1$ \cite{smixing}.
However, because the other solution may exist in which flavor mixing
can be achieved only partially \cite{smixing}, and because the
evolutions of CMB anisotropies are completely independent from BBN
physics, it will still be important to investigate 
how much the sensitivity of CMB anisotropies for $\xi$ can be improved by
including the effect of $\xi$ on the evolution of gravitational waves
studied here.

%--------------------------------------------------------------------------

\acknowledgments{ JY thank Galileo Galilei Institute for Theoretical
Physics for the hospitality and the INFN for partial support during
the completion of this work.
This work was partially supported by JSPS Grant-in-Aid
for Scientific Research Nos. 1809940(KI), 18740157(MY), and
16340076(JY). M.Y. is supported in part by the project of the Research
Institute of Aoyama Gakuin University. }

\end{document}